\documentclass[11pt,letterpaper]{article}
\usepackage{array, latexsym, graphicx,amsfonts,picture,color,amsmath,verbatim,blkarray,setspace,subfig,multirow,array,url,epsf,epsfig,breqn,listings,txfonts,textcomp}
\begin{document}

\newtheorem{theorem}{Theorem}[section]
\newtheorem{ex}[theorem]{Example}

\title{Quantum Resistant Random Linear Code Based Public Key Encryption Scheme RLCE}
\author{Yongge Wang\\ Department of SIS, UNC Charlotte, USA. \\
{yongge.wang@uncc.edu}}

\maketitle

\begin{abstract}
Lattice based encryption schemes and linear code based encryption schemes
have received extensive attention in recent years since they have been 
considered as post-quantum candidate encryption schemes. 
Though LLL reduction algorithm has been one of the major cryptanalysis techniques
for lattice based cryptographic systems, key recovery cryptanalysis techniques for linear code
based cryptographic systems are generally scheme specific. In recent years,
several important techniques such as  Sidelnikov-Shestakov attack, filtration attacks,
and algebraic attacks 
have been developed to crypt-analyze linear code based encryption schemes. 
Though most of these cryptanalysis techniques
are relatively new, they prove to be very powerful and many systems have been broken
using them. Thus it is important to  design linear code 
based cryptographic systems that are immune against these attacks.
This paper proposes linear code based encryption scheme RLCE which shares
many characteristics with random linear codes. Our analysis shows
that the scheme RLCE is secure against existing attacks and we hope that 
the security of the RLCE scheme is equivalent to the hardness of decoding random linear codes.
Example parameters for different security levels are recommended for the scheme RLCE.
\end{abstract}



\section{Introduction}
\label{introsec}
With rapid development for quantum computing techniques, 
our society is concerned with the security of current 
Public Key Infrastructures (PKI)
which are fundamental for Internet services.
The core components for current PKI infrastructures are based on public 
cryptographic techniques such as RSA and DSA.
However, it has been shown that these public key cryptographic techniques 
could be broken by quantum computers. Thus it is urgent to develop 
public key cryptographic systems that are secure against quantum computing.

Since McEliece encryption scheme  \cite{mceliece1978public} was 
introduced more than thirty years ago, it has withstood many attacks 
and still remains unbroken for general cases. It has been considered 
as one of the candidates for post-quantum cryptography since it is 
immune to existing quantum computer algorithm attacks. The original 
McEliece  cryptographic system is based on binary Goppa codes. Several variants 
have been introduced to replace Goppa codes  in the McEliece encryption scheme. 
For instance, Niederreiter \cite{niederreiter1986knapsack}
proposed the use of generalized Reed-Solomon codes and later,
Berger and Loidreau \cite{berger2005mask} proposed the use of sub-codes of 
generalized Reed-Solomon codes.
Sidelnikov \cite{sidelnikov1994public} proposed the use of Reed-Muller codes, 
Janwa and Moreno \cite{janwa1996mceliece} proposed the use of 
algebraic geometry codes,  Baldi et al \cite{baldi2008new}
proposed the use of LDPC codes, Misoczki et al \cite{misoczki2013mdpc} proposed the use of 
MDPC codes, 
L{\"o}ndahl and Johansson \cite{londahl2012new} proposed the use of convolutional codes,
and Berger et al \cite{berger2009reducing} and
Misoczki-Barreto \cite{misoczki2009compact} proposed quasi-cyclic 
and quasi-dyadic structure based 
compact variants of McEliece encryption schemes.
Most of them have been broken though MDPC/LDPC code based  
McEliece encryption scheme \cite{baldi2008new,misoczki2013mdpc} 
and the original binary Goppa code based McEliece encryption scheme 
are still considered secure.

Goppa code based McEliece encryption scheme is hard to attack 
since Goppa codes share many characteristics with random codes.  
Motivated by Faugere et al's \cite{fau2010algebraicae} algebraic attacks
against quasi-cyclic and quasi-dyadic structure based 
compact variants of McEliece encryption schemes,
Faugere et al \cite{faugere2013distinguisher} designed an efficient
algorithm to distinguish a random code from a high rate Goppa code.
M{\'a}rquez-Corbella and Pellikaan \cite{marquez2012error} simplified the distinguisher
in \cite{faugere2013distinguisher} using Schur component-wise product of codes.

Sidelnikov and Shestakov \cite{sidelnikov1992insecurity} showed, for the
generalized Reed-Solomon code based  McEliece encryption scheme, one can efficiently
recover the private parameters for the generalized Reed-Solomon code from the public key.
Using  component-wise product of codes and techniques from \cite{sidelnikov1992insecurity}, 
Wieschebrink \cite{wieschebrink2010cryptanalysis} showed that 
Berger and Loidreau's proposal \cite{berger2005mask}
could be broken efficiently also. Couvreur et al \cite{couvreur2013distinguisher} 
proposed a general distinguisher based filtration technique to recover keys
for generalized Reed-Solomon code based McEliece scheme and 
Couvreur, M{\'a}rquez-Corbella, and Pellikaan \cite{couvreur2014isit}
used filtration attacks to break Janwa and Moreno's  \cite{janwa1996mceliece} 
algebraic geometry code based McEliece encryption scheme. 
The filtration attack was recently used by Couvreur et al \cite{couvreur2014polynomial} 
and Faugere et al \cite{faugere2014algebraic}
to attack  Bernstein et al's  \cite{bernstein2011wild} wild Goppa code based McEliece scheme.

General Goppa code based McEliece schemes are still immune from these attacks. 
However, based on the new development of cryptanalysis techniques against 
linear code based cryptographic systems in the recent years, it is important to 
systematically design random linear code based cryptographic systems
defeating these attacks. Motivated by this observation, this paper presents a systematic
approach of designing public key encryption schemes using any linear code.
For example, we can even use Reed-Solomon codes to design
McEliece encryption scheme while it is insecure to use Reed-Solomon codes
in the original McEliece scheme. Since our design of linear code based encryption 
scheme embeds randomness in each column of the generator matrix, it is 
expected that, without the corresponding private key, these codes are 
as hard as random linear codes for decoding.

The most powerful message recovery attacks (not key recovery attacks) on McEliece cryptosystem
is the information-set decoding attack which was introduced by Prange \cite{prange1962use}.
In an information-set decoding approach, one finds a set of coordinates 
of a received ciphertext which are error-free and that the restriction of the code's
generator matrix to these positions is invertible. The original message can then 
be computed by multiplying the ciphertext with the inverse of the sub-matrix.
Improvements of the information-set decoding attack have been 
proposed by Lee-Brickell \cite{lee1988observation}, Leon \cite{leon1988probabilistic}, 
Stern \cite{stern1989method}, May-Meurer-Thomae \cite{may2011decoding},
Becker-Joux-May-Meurer \cite{becker2012decoding},
and May-Ozerov \cite{may2015computing}.
Bernstein, Lange, and Peters \cite{bernstein2008attacking}
presented an exact complexity analysis on information-set 
decoding attack against McEliece cryptosystem. The attacks in \cite{becker2012decoding,bernstein2008attacking,lee1988observation,leon1988probabilistic,may2011decoding,may2015computing,stern1989method}
are against binary linear codes and are not applicable 
when the underlying field is $GF(p^m)$ for a prime $p$. 
Peters \cite{peters2010information} presented an exact complexity analysis on information-set 
decoding attack against McEliece cryptosystem over $GF(p^m)$. 
These information-set decoding techniques (in particular, the exact complexity analysis 
in \cite{bernstein2008attacking,peters2010information}) 
are used to select example parameters for RLCE scheme in Section \ref{pracsec}. 

Unless specified otherwise, we will use $q=2^m$ or $q=p^m$ for a prime $p$ and our discussion 
are based on the field $GF(q)$ through out this paper. Bold face letters 
such as ${\bf a, b, e, f, g}$ are used to denote row or column vectors
over $GF(q)$. It should be clear from the context whether a specific bold
face letter represents a row vector or a column vector.

\section{Goppa codes and  McEliece Public Key Encryption scheme}
\label{goppasec}
In this section, we briefly review Goppa codes and  McEliece scheme.
For given parameters $q, n\le q$, and $t$, let $g(x)$ be a polynomial 
of degree $t$ over  $GF(q)$. Assume that
$g(x)$ has no  multiple zero roots and 
$\alpha_0, \cdots, \alpha_{n-1}\in GF(q)$ be pairwise distinct which are not root of $g(x)$.
The following subspace ${\cal C}_{Goppa}(g)$ defines the code words of an $[n,k,d]$ binary Goppa code
where $d\ge 2t+1$. This binary Goppa code ${\cal C}_{Goppa}(g)$ has dimension $k\ge n-tm$ and 
corrects $t$ errors.
$${\cal C}_{Goppa}(g)=\left\{
c\in \{0,1\}^n: \sum_{i=0}^{n-1}\frac{c_i}{x-\alpha_i}\equiv 0 \mbox{ mod } g(x) 
\right\}.
$$
Furthermore, if $g(x)$ is irreducible, then ${\cal C}_{Goppa}(g)$ is called an irreducible Goppa code.
The parity check matrix $H$ for the Goppa codes looks as follows:
\begin{equation}
\label{parityhv}
V_t({\bf x,y})=
\left[
\begin{array}{cccc}
1 & 1 & \cdots & 1\\
\alpha_0^1&\alpha_1^1&\cdots & \alpha_{n-1}^1\\
\ldots & \ldots &\ddots & \ldots\\
\alpha_0^t&\alpha_1^t&\cdots & \alpha^t_{n-1}\\
\end{array}
\right]
\left[
\begin{array}{cccc}
\frac{1}{g(\alpha_0)}&&&\\
&&\ddots&\\
&&&\frac{1}{g(\alpha_{n-1})}\\
\end{array}
\right]
\end{equation}
where ${\bf x}=[\alpha_0, \ldots, \alpha_{n-1}]$ and 
${\bf y}=\left[\frac{1}{g(\alpha_0)}, \ldots, \frac{1}{g(\alpha_{n-1})}\right]$.

The McEliece scheme  \cite{mceliece1978public} is described as follows.
For the given parameters $n$ and $t$, choose a binary Goppa code based on an irreducible 
polynomial $g(x)$ of degree $t$. Let $G_s$ be the $k\times n$ generator 
matrix for the Goppa code. Select a random dense $k\times k$ nonsingular 
matrix $S$ and a random $n\times n$ permutation matrix $P$. Note that the permutation 
matrix $P$ is required only if the support $\alpha_0, \cdots, \alpha_{n-1}$ is known to the public.
Then the public key is $G=SG_sP$ which generates a linear code 
with the same rate and minimum distance as the code generated by $G_s$.
The private key is $G_s$.

\noindent
{\em Encryption}. For a $k$-bit message block ${\bf m}$, choose a random 
row vector ${\bf e}$ of length $n$ and weight $t$. 
Compute the cipher text ${\bf y}={\bf m}G+{\bf e}$

\noindent
{\em Decryption}. For a received ciphertext ${\bf y}$, first compute 
${\bf y}'={\bf y}P^{-1}$. Next use an error-correction algorithm to recover 
${\bf m}'={\bf m}S$ and compute the message ${\bf m}$ as
${\bf m}={\bf m}'S^{-1}$.

\section{Random linear code based encryption scheme RLCE}
The protocol for the Random Linear Code based Encryption scheme RLCE
proceeds as follows:

\noindent
{\em Key Setup.}
Let $n$, $k, d,t>0$, and $r\ge 1$ be given parameters such that 
$n-k+1\ge d\ge 2t+1$.
Let $G_s=[{\bf g}_0, \cdots, {\bf g}_{n-1}]$ be a $k\times n$ 
generator matrix for an $[n,k,d]$ linear code such that there is an efficient 
decoding algorithm to correct at least $t$ errors 
for this linear code given by $G_s$. 
\begin{enumerate}
\item Let $C_0, C_1, \cdots,C_{n-1} \in GF(q)^{k\times r}$ be $k\times r$  matrices drawn uniformly at random and let 
\begin{equation}
G_1=[{\bf g}_0, C_0,  {\bf g}_1, C_1 \cdots,  {\bf g}_{n-1},C_{n-1}]
\end{equation} 
be the $k\times n(r+1)$ matrix obtained by inserting the random
matrices $C_i$ into $G_s$.
 \item Let $A_0, \cdots, A_{n-1}\in GF(q)^{(r+1)\times (r+1)}$ be
dense nonsingular $(r+1)\times (r+1)$ matrices chosen uniformly at random and let 
\begin{equation}
\label{defineA}
A=\left[
\begin{array}{cccc}
A_0& & &\\
&A_1&&\\
&&\ddots&\\
&&&A_{n-1}\\
\end{array}
\right]
\end{equation}
be an  $n(r+1)\times n(r+1)$ nonsingular matrix.
\item Let $S$ be a random dense $k\times k$ nonsingular matrix and  $P$ be an $n(r+1)\times n(r+1)$ permutation matrix.
\item The public key is the $k\times n(r+1)$ matrix $G=SG_1AP$ and the private key is 
$(S, G_s, P, A)$.
\end{enumerate}

\noindent
{\em Encryption}. For a row vector message ${\bf m}\in GF(q)^{k}$, choose 
a random row vector ${\bf e}=[e_0, \ldots, e_{n(r+1)-1}]\in GF(q)^{n(r+1)}$ such that the Hamming weight 
of ${\bf e}$ is at most $t$. The cipher text is ${\bf y}={\bf m}G+{\bf e}$.

\noindent
{\em Decryption}.   For a received cipher text ${\bf y}=[y_0, \ldots,  y_{n(r+1)-1}]$, compute 
$${\bf y}P^{-1}A^{-1}=[y'_0, \ldots, y'_{n(r+1)-1}]={\bf m}SG_1 + {\bf e}P^{-1}A^{-1}$$
where 
\begin{equation}
\label{Ainverse}
A^{-1}=\left[
\begin{array}{cccc}
A_0^{-1}& & &\\
&A_1^{-1}&&\\
&&\ddots&\\
&&&A_{n-1}^{-1}\\
\end{array}
\right]
\end{equation}
Let ${\bf y}'=[y'_0, y'_{r+1}, \cdots, y'_{(n-1)(r+1)}]$ be the row vector of length $n$
selected from the length $n(r+1)$ row vector ${\bf y}P^{-1}A^{-1}$. 
Then ${\bf y}'={\bf m}SG_s+ {\bf e}'$ for some error vector ${\bf e}'\in GF(q)^n$.
Let ${\bf e}''={\bf e}P^{-1}=[e''_0, \cdots, e''_{n(r+1)-1}]$ and 
${\bf e}''_i=[e''_{i(r+1)},  \ldots, e''_{i(r+1)+r}]$ be a sub-vector of ${\bf e}''$
for  $ i\le n-1$. Then ${\bf e}'[i]$ is the first element of ${\bf e}''_iA_i^{-1}$.  Thus ${\bf e}'[i]\not=0$ only if 
${\bf e}''_i$ is non-zero. Since there are at most $t$ non-zero sub-vectors ${\bf e}''_i$, 
the Hamming weight of ${\bf e}'\in GF(q)^{n}$ is at most $t$.
Using the efficient decoding algorithm, one can compute ${\bf m}'={\bf m}S$ 
and ${\bf m}={\bf m}'S^{-1}$. Finally, calculate the Hamming weight 
$w={\tt weight}({\bf y}-{\bf m}G)$. If $w\le t$ then output ${\bf m}$ as the 
decrypted plaintext. Otherwise, output error.

\noindent
{\bf Comment 1}. In the design of RLCE scheme, the permutation matrix $P$ has two purposes.
The first purpose is to hide the supports of the underlying encoding scheme generator matrix
(this is necessary if the supports of the underlying encoding scheme are unknown).
The second purpose is to hide the positions and combinations of the column vectors 
${\bf g}_i$ and $C_i$.

\noindent
{\bf Comment 2}. In the RLCE decryption process, one checks whether 
the Hamming weight $w={\tt weight}({\bf y}-{\bf m}G)$ is smaller than $t$. This step 
is used to defeat chosen ciphertext attacks (CCA). In a CCA atack, an adversary gives
a random vector ${\bf y}=[y_0, \ldots,  y_{n(r+1)-1}]$ (which is not 
a valid ciphertext) to the decryption oracle to learn a decrypted value.
This decrypted value could be used to obtain certain information
about the private generator matrix $G_s$  (see Section \ref{ccasec} for details).
Alternatively, one may use an appropriate padding scheme to pad a message before encryption.
Then it is sufficient for the decryption process to verify whether the decrypted message 
has the correct padding strings to defeat the CCA attacks.

\section{Robustness of RLCE codes against existing attacks}
\label{revisedsec}
\subsection{Randomness of generator matrix columns}
\label{randomnsfer}
We first use the following theorem to show that any single column of the underlying 
generator matrix $G_s$ could be completely randomized in a RLCE public key $G$.
\begin{theorem}
\label{thmlinear}
Let $G_s=[{\bf g}_0, \cdots, {\bf g}_{n-1}]\in GF(q)^{k\times (n-1)}$ be a linear code 
generator matrix. 
For any randomly chosen full rank $k\times (r+1)$  matrix $R_0\in GF(q)^{k\times (r+1)}$, 
there exists a $k\times k$ nonsingular matrix $S$, a $(r+1)\times (r+1)$
matrix $A_0$, and a $k\times r$ matrix $C_0\in GF(q)^{k\times r}$ such that 
\begin{equation}
R_0=S[{\bf g}_0, C_0] A_0
\end{equation}
\end{theorem}

{\em Proof.} By the fundamental properties of matrix equivalence, for two $m\times n$
matrices $A$,  $B$ of the same rank, there exist invertible $m\times m$ matrix $P$ and $n\times n$
invertible matrix $Q$ such that $A=PBQ$. The theorem could be proved using this property and 
the details are omitted here. 
\hfill$\Box$

Let $R=[R_0, \ldots, R_{n-1}]\in GF(q)^{k\times n(r+1)}$
be a fixed random linear code generator matrix.
Theorem \ref{thmlinear} shows that for any generator matrix $G_s$ (e.g.,
a Reed-Solomon code generator matrix), 
we can choose matrices $S$ and $A_0$ so that the first $r+1$
columns of the RLCE scheme public key $G$ (constructed from $G_s$) 
are identical to $R_0$. However, we cannot use Theorem \ref{thmlinear} to continue
the process of choosing $A_1, \ldots, A_{n-1}$ to obtain $G=R$ 
since $S$ is fixed after $A_0$ is chosen. Indeed, it is straightforward to show
that one can use Theorem \ref{thmlinear} to continue
the process of choosing $A_1, \ldots, A_{n-1}$ to obtain $G=R$ 
if and only if there exists a $k\times k$ nonsingular matrix $S$
such that, for each $i\le n-1$, the vector $S{\bf g}_i$ lies in the linear space
generated by the column vectors of $R_i$. A corollary of this observation 
is that if $R_i$ generates the full $k$ dimensional space, then 
each linear code could have any random matrix as its RLCE public key.

\begin{theorem}
\label{fulatre}
Let $R=[R_0, \ldots, R_{n-1}]\in GF(q)^{k\times n(r+1)}$ and 
$G_s=[{\bf g}_0,\cdots, {\bf g}_{n-1}]\in GF(q)^{k\times n}$ be two fixed MDS
linear code generator matrices. If $r+1\ge k$, then there exist
$A_0,\cdots, A_{n-1}\in GF(q)^{(r+1)\times (r+1)}$ and $C_0,\cdots, C_{n-1}\in GF (q)^{k\times r}$ 
such that $R =[{\bf g}_0, C_0, \cdots, {\bf g}_{n-1}, C_{n-1}]A$ where $A$ 
is in the format of (\ref{defineA}).
\end{theorem}

{\em Proof.} Without loss of generality, we may assume that 
$r=k-1$. For each $0\le i\le n-1$, choose a random matrix $C_i\in GF (q)^{k\times r}$ 
such that $G_i=[{\bf g}_i, C_i]$ is an $k\times k$ invertible matrix.
Let $A=G_i^{-1}R_i$. Then the theorem is proved.
\hfill$\Box$

Theorem \ref{fulatre} shows that in the RLCE scheme, we must have $r< k-1$. Otherwise, 
for a given public key $G\in GF(q)^{k\times n(r+1)}$, the adversary can choose a 
Reed-Solomon code generator matrix $G_s\in GF(q)^{k\times n}$ and compute 
$A_0, \cdots,A_{n-1}\in GF(q)^{r\times r}$ and $C_0,\cdots ,C_{n-1}\in GF(q)^{k\times r}$ 
such that $G =[{\bf g}_0, C_0, \cdots, {\bf g}_{n-1}, C_{n-1}]A$. In other words, 
the adversary can use the decryption algorithm corresponding to the generator matrix $G_s$ 
to break the RLCE scheme

Theorem \ref{fulatre} also implies an efficient decryption algorithm for random 
$[n, k]$ linear codes with sufficiently small $t$ of errors. 
Specifically, for an $[n,k]$ linear code with generator matrix $R\in GF(q)^{k\times n}$, 
if $t\le \frac{n-k^2}{2k}$, then one can divide $R$ into $m=2t+k$ blocks $R = [R_0,\cdots, R_{m-1}]$. 
Theorem \ref{fulatre}  can then be applied to construct an equivalent $[m, k]$ Reed-Solomon code 
with generator matrix $G_s\in GF(q)^{k\times m}$. Thus it is sufficient to decrypt the equivalent 
Reed-Solomon code instead of the original random linear code. For McEliece based encryption 
scheme, Bernstein, Lange, and Peters \cite{bernstein2008attacking} recommends 
the use of 0.75 ($= k/n$) as the code rate. 
Thus Theorem \ref{fulatre}  has no threat on these schemes.

For $t\le \frac{n-k^2}{2k}$, the adversary is guaranteed to succeed in breaking the system. 
Since multiple errors might be located within the same block $R_i$ with certain probability, 
for a given $t$ that is slightly larger than $\frac{n-k^2}{2k}$, the adversary still has a good 
chance to break the system using the above approach. It is recommended that $t$ is
significantly larger than $\frac{n-k^2}{2k}$. For the  RLCE scheme, this means that $r$ 
should be significantly smaller than $k$. This is normally true since $k$ is 
very larger for secure RLCE schemes. 

In  following sections, we list heuristic and experimental evidences
that the RLCE public key $G$ shares the properties of random linear codes. 
Thus the security of the RLCE scheme is believed to be 
equivalent to decoding a random linear code which is {\bf NP}-hard.

\subsection{Chosen ciphertext attacks (CCA)} 
\label{ccasec} 
In this section, we show that certain information about the private generator
matrix $G_s$ is leaked if the decryption process does neither include 
padding scheme validation nor include ciphertext correctness validation.
However, it is not clear whether this kind of information leakage would
help the adversary to break the RLCE encryption scheme. We illustrate this using
the parameter $r=1$.

Assume that $G_1=[{\bf g}_{0}, {\bf r}_{0}, {\bf g}_{1}, {\bf r}_{1}, \cdots,  {\bf g}_{n-1},{\bf r}_{n-1}]$ 
and $G=SG_1AP$.  The adversary chooses a random vector 
${\bf y}=[y_0, \ldots,  y_{2n-1}]\in GF(q)^{2n-1}$
and gives it to the decryption oracle which outputs a vector ${\bf x}\in GF(q)^{k}$.
Let ${\bf y}P^{-1}A^{-1}=[y'_0, \ldots, y'_{2n-1}]$ and 
$A_i=\left[
\begin{array}{cc}
a_{i,00}& a_{i,01}\\
a_{i,10}&a_{i,11}\\
\end{array}
\right]$.
Then we have
\begin{equation}
\label{ai0111ratio}
\begin{array}{lll}
{\bf x}G-{\bf y}&=&{\bf x}S[{\bf g}_{0}, {\bf r}_{0}, {\bf g}_{1}, {\bf r}_{1}, \cdots,  {\bf g}_{n-1},{\bf r}_{n-1}]AP-{\bf y}\\
&=&[\cdots, {\bf x}S[{\bf g}_{i}, {\bf r}_{i}]A_i, \cdots]P-{\bf y}\\
&=&[\cdots, [y'_{2i},  {\bf x}S{\bf r}_{i}]A_i, \cdots]P+{\bf e}-{\bf y}\\
&=&[\cdots, [y'_{2i},  y'_{2i+1}]A_i, \cdots]P+ [\cdots, [0,  {\bf x}S{\bf r}_{i}-y'_{2i+1}]A_i, \cdots]P+{\bf e}-{\bf y}\\
&=&{\bf y}+ [\cdots, [0,  {\bf x}S{\bf r}_{i}-y'_{2i+1}]A_i, \cdots]P+{\bf e}-{\bf y}\\
&=&[\cdots, [({\bf x}S{\bf r}_{i}-y'_{2i+1})a_{i,10}, ({\bf x}S{\bf r}_{i}-y'_{2i+1})a_{i,11}], \cdots]P+{\bf e}\\
\end{array}
\end{equation}
where ${\bf e}$ is a row vector of Hamming weight at most $t$.
From the identity (\ref{ai0111ratio}), one can calculate a list of potential values 
for $c_i=a_{i,10}/a_{i,11}$. The size of this list is ${2n \choose 2}$.
For each value in this list, one obtains the corresponding two
column vectors $[{\bf f}_0,{\bf f}_1]=S[{\bf g}_i, {\bf r}_i]A_i$ from the public key $G$. 
Then we have 
\begin{equation}
\label{rerefd}
[{\bf f}_0,{\bf f}_1]\left[\begin{array}{cc}
1& 0\\
-c_i& 1\\
\end{array}
\right]=
S[{\bf g}_i, {\bf r}_i]\left[\begin{array}{cc}
a_{i,00}& a_{i,01}\\
c_ia_{i,11}&a_{i,11}\\
\end{array}
\right]\left[\begin{array}{cc}
1& 0\\
-c_i& 1\\
\end{array}
\right]
=S[{\bf g}_i, {\bf r}_i]\left[\begin{array}{cc}
a_{i,00}-c_ia_{i,01}& a_{i,01}\\
0&a_{i,11}\\
\end{array}
\right]
\end{equation}
That is, ${\bf f}_0-c_i{\bf f}_1=(a_{i,00}-c_ia_{i,01})S{\bf g}_i$. 
Thus, for each candidate permutation matrix $P$, one can calculate
a matrix $SG_sB$ where $B=\mbox{diag}[a_{0,00}-c_0a_{0,01}, \cdots, a_{n-1,00}-c_{n-1}a_{n-1,01}]$
is an $n\times n$ diagonal matrix with unknown diagonal elements
$a_{0,00}-c_0a_{0,01}, \cdots$, and $a_{n-1,00}-c_{n-1}a_{n-1,01}$.

On the other hand, for each ciphertext ${\bf y}=[y_0, \ldots,  y_{2n-1}]\in GF(q)^{2n-1}$,
let ${\bf y}P^{-1}=[z_0, z_1, \cdots, z_{2n-1}]$.
The codeword corresponding to the secret generator matrix $SG_s$ is 
$[y'_0, y'_2, \ldots, y'_{2n-2}]$ where
${\bf y}P^{-1}A^{-1}=[y'_0, \ldots, y'_{2n-1}]$. 
By the fact that
$$[y'_{2i},y'_{2i+1}]= [z_{2i}, z_{2i+1}]A_i^{-1} 
=\frac{1}{|A_i|}[z_{2i}, z_{2i+1}]\left[\begin{array}{cc}
a_{i,11}& -a_{i,01}\\
-c_ia_{i,11}&a_{i,00}\\
\end{array}
\right],
$$
we have $y'_{2i}=\frac{a_{i,11}}{|A_i|}(z_{2i}-c_iz_{2i+1})$.
For each candidate permutation matrix $P$, one first 
chooses $k$ independent messages ${\bf x}_0, \cdots, {\bf x}_{k-1}$
and calculates the corresponding $k$ independent ciphertexts 
${\bf y}_0, \cdots, {\bf y}_{k-1}$. Using $P$ and the above mentioned
technique, one obtain a generator matrix 
$G_a=S'G_s\mbox{diag}\left[\frac{a_{0,11}}{|A_0|}, \cdots, \frac{a_{n-1,11}}{|A_{n-1}|}\right]$.
Thus in order to decode a ciphertext ${\bf y}$, it is sufficient to decode 
the error correcting code given by the generator matrix  $G_a$. 
This task becomes feasible for certain codes. For example, this task is equivalent 
to the problem of attacking a generalized Reed-Solomon code based 
McElience encryption scheme if $G_s$ generates a generalized Reed-Solomon code.

In order for the attacks in the preceding paragraphs to work, 
the adversary needs to have the knowledge of the permutation matrix $P$. Since 
the number of candidate permutation matrices $P$ is huge, 
this kind of attacks is still infeasible in practice.

\subsection{Niederreiter's scheme and Sidelnikov-Shestakov's attack} 
Sidelnikov and Shestakov's cryptanalysis technique
\cite{sidelnikov1992insecurity} was used to 
analyze Niederreiter's scheme which is based on generalized Reed-Solomon codes.
Let $\alpha=(\alpha_0, \ldots, \alpha_{n-1})$ be $n$ distinct elements of $GF(q)$ and let 
$v=(v_0,\ldots, v_{n-1})$ be nonzero (not necessarily distinct) elements of $GF(q)$. The 
generalized Reed-Solomon (GRS) code of dimension $k$, denoted by $GRS_k(\alpha,v)$, 
is defined by the following subspace.
$$GRS_k(\alpha,v)=
\left\{ (v_0f(\alpha_0), \ldots, v_{n-1}f(\alpha_{n-1})): f(x)\in GF(q)[x]_k)
\right\}
$$
where  $GF(q)[x]_k$ is the set of polynomials in  $GF(q)[x]$ of degree less than $k$.
$GF(q)[x]_k$ is a vector space of dimension $k$ over $GF(q)$.
For each code word  $c=(v_0f(\alpha_0), \ldots, v_{n-1}f(\alpha_{n-1}))$, 
$f(x)=f_0+f_1x+\ldots+f_{k-1}x^{k-1}$ is called the associate polynomial of the code word $c$
that encodes the message $(f_0, \ldots, f_{k-1})$. 
$GRS_k(\alpha,v)$ is an $[n,k,d]$ MDS code where $d=n-k+1$. 

Niederreiter's scheme  \cite{niederreiter1986knapsack} replaces
the binary Goppa codes in McEliece scheme using GRS codes. 
The first attack on Niederreiter scheme is presented by 
Sidelnikov and Shestakov \cite{sidelnikov1992insecurity}. 
Wieschebrink \cite{wieschebrinkreisedMce} revised  Niederreiter's scheme by inserting 
random column vectors into random positions of $G_s$ before obtaining the public key $G$.
Couvreur et al \cite{couvreur2013distinguisher} showed that 
Wieschebrink's revised scheme is insecure under the product code attacks.

Berger and Loidreau \cite{berger2005mask} recommend the use of sub codes of Niederreiter's scheme
to avoid Sidelnikov and Shestakov's attack. Specifically, in Berger and Loidreau's scheme, 
one uses a random $(k-l)\times k$ matrix $S'$ of rank $k-l$  instead of the $k\times k$
matrix $S$ to compute the public key $G=S'G_s$.  

For smaller values of $l$, Wieschebrink \cite{wieschebrink2010cryptanalysis} shows that 
a private key $(\alpha,v)$ for Berger and Loidreau scheme \cite{berger2005mask}  
could be recovered using Sidelnikov-Shestakov algorithm. For larger values 
of $l$,  Wieschebrink used Schur product code to recover the secret values for 
Berger-Loidreau scheme.
Let $G=S G_s$ be the $(k-l)\times n$ public key generator matrix for Berger-Loidreau scheme, 
${\bf r}_0, \cdots, {\bf r}_{k-l-1}$ be the rows of $G$, and $f_0, \cdots, f_{k-l-1}$ be 
the associated polynomials to those rows. For two row vectors ${\bf a, b}\in GF(q)^n$,
the component wise product ${\bf a}* {\bf b}\in GF(q)^n$ is defined as
\begin{equation}
\label{stardef}
{\bf a}*{\bf b}=(a_0b_0, \cdots, a_{n-1}b_{n-1})
\end{equation}
By the definition in (\ref{stardef}), it is straightforward to observe that
\begin{equation}
{\bf r}_i*{\bf r}_j=(v_0^2f_i(\alpha_0)f_j(\alpha_0), \cdots, v_{n-1}^2f_i(\alpha_{n-1})f_j(\alpha_{n-1})).
\end{equation}
For $2k-1\le n-2$, if the code generated by ${\bf r}_i*{\bf r}_j$ equals $GRS_{2k-1}(\alpha,{\bf v}')$ 
for ${\bf v}'=(v_0^2,\cdots, v_{n-1}^2)$, then the 
Sidelnikov-Shestakov algorithm could be used to recover the values $\alpha$ and $v$. 
For $2k-1\le n-2$, if the code generated by ${\bf r}_i*{\bf r}_j$ does not equal $GRS_{2k-1}(n,{\bf v}')$,
then the attack fails. Wieschebrink claimed that the probability that the attack fails is 
very small.
For the case of $2k-1>n-2$, Wieschebrink 
applied Sidelnikov-Shestakov algorithm on the component wise product code of 
a shortened code of the original $GRS_k(\alpha,{\bf v})$.

The crucial step in Sidelnikov and Shestakov attack is to use the 
echelon form $E(G)=[I | G']$ of the public key to get minimum 
weight codewords that are co-related to each other supports. 
In the encryption scheme RLCE,
each column of the public key $G$ contains mixed 
randomness. Thus the  echelon form $E(G)=[I | G']$ 
obtained from the public key $G$ could not be 
used to build any useful equation system. In other words, 
it is expected that Sidelnikov and Shestakov attack
does not work against the RLCE scheme.

\subsection{Filtration attacks}
Using  distinguisher techniques \cite{faugere2013distinguisher}, Couvreur et al.  \cite{couvreur2013distinguisher} designed a filtration technique to attack 
GRS code based McEliece scheme.
The filtration technique was further developed  by 
Couvreur et al \cite{couvreur2014polynomial} to attack 
wild Goppa code based McEliece scheme. In the following, 
we briefly review the filtration attack in \cite{couvreur2014polynomial}. 
For two codes ${\cal C}_1$ and ${\cal C}_2$ of length $n$, the star product code
${\cal C}_1*{\cal C}_2$ is the vector space spanned by ${\bf a*b}$ for all pairs 
$({\bf a,b})\in {\cal C}_1\times {\cal C}_2$ where  ${\bf a*b}$  is defined in  (\ref{stardef}).
For  ${\cal C}_1={\cal C}_2$,  ${\cal C}_1*{\cal C}_1$
is called the square code of ${\cal C}_1$.  It is showed in 
\cite{couvreur2014polynomial} that 
\begin{equation}
\label{bequagd}
\dim{\cal C}_1\times {\cal C}_2\le \left\{n, \dim{\cal C}_1\dim{\cal C}_2
-{\dim({\cal C}_1\cap{\cal C}_1)\choose 2}  \right\}.
\end{equation}
Furthermore, the equality in (\ref{bequagd}) is attained for 
most randomly selected codes ${\cal C}_1$ and ${\cal C}_2$ 
of a given length and dimension. Note that for ${\cal C}={\cal C}_1={\cal C}_2$
and $\dim{\cal C}=k$, the equation (\ref{bequagd}) becomes
$\dim{\cal C}^{*2}\le \min\left\{n, {k+1\choose 2}\right\}$.

Couvreur et al \cite{couvreur2014polynomial} showed that the square code of an  alternant code of extension degree 
$2$  may have an unusually low dimension when its actual rate is larger than its designed rate. Specifically, Couvreur et al created a family of nested codes (called a filtration) 
defined as follows, for any $a\in\{0, \cdots, n-1\}$:
\begin{equation}
\label{nested}
{\cal C}^a(0)\supseteq {\cal C}^a(1)\supseteq \cdots \supseteq {\cal C}^a(q+1).
\end{equation}
Roughly speaking, ${\cal C}^a(j)$  consists in the codewords of ${\cal C}$ which correspond
to polynomials which have a zero of order $j$ at position $a$. 
The first two elements of this filtration are just punctured and shortened versions of ${\cal C}$ 
and the rest of them can be computed from ${\cal C}$ by computing star products and solving linear systems. 
The support values $\alpha_0, \cdots, \alpha_{n-1}$ (the private key) 
for the Goppa code could be recovered using this nested family of codes efficiently. 

The crucial part of the filtration technique is the efficient algorithm to 
compute the nested family of codes in (\ref{nested}). For our RLCE scheme, 
the public key generator matrix $G$ contains random columns.
Thus linear equations constructed in Couvreur et al \cite{couvreur2014polynomial} 
could not be solved and the nested family (\ref{nested}) could not be computed 
correctly. Furthermore, the important characteristics for a code ${\cal C}$
to be vulnerable is that one can find a related code ${\cal C}_1$ of dimension $k$
such that the dimension of the square code of ${\cal C}_1$ has a dimension significantly less
than $\min\left\{n, {k+1\choose 2}\right\}$.

To get experimental evidence that RLCE codes share similarity with random
linear codes with respect to the above mentioned filtration  attacks, 
we carried out several experiments using  Shoup's NTL library \cite{shoup2001ntl}.
The source code for our experiments is available at \cite{wangrlcesoft}.
In the experiments, we used Reed-Solomon codes over $GF(2^{10})$.
The RLCE parameters are chosen as
the 80-bit security parameter $n=560$, $k=380$, $t=90$, and $ r=1$
(see Section \ref{pracsec} for details).
For each given $380\times 560$ generator matrix $G_s$ of Reed-Solomon code,
we selected another random $380\times 560$ matrix $C\in GF(2^{10})^{380\times 560}$
and selected $2\times 2$ matrices $A_0, \ldots, A_{559}$. Each column ${\bf c}_i$
in $C$ is inserted in $G_s$ after the column ${\bf g}_i$. The extended generator matrix 
is multiplied by $A=\mbox{diag}[ A_0, \ldots, A_{559}]$ from the right hand side
to obtain the public key matrix $G\in GF(2^{10})^{380\times 1120}$. 
For each $i=0, \cdots, 1119$, the matrix $G_{i}$ is used to compute the product
code, where $G_{i}$ is obtained from $G$ by deleting the $i$th column vector. 
In our experiments, all of these product codes have dimension
$1119$. We repeated the above experiments 100 times for 100 distinct Reed-Solomon generator matrices
and the results remained the same. Since $\min\left\{1119, {381\choose 2}\right\}=1119$, the experimental results 
meet our expectation that RLCE behaves like a random linear code.
We did the same experiments for the dual code of the above code. That is,
for a $180\times 560$ generator matrix $G_s$ of the dual code, the same procedure has
been taken. In this time, after deleting one column from the resulting public key matrix,
the product code always had dimension 1119 which is the expected dimension for a random linear code.
In an early draft of this paper, we used Maple 2015 to carry out the experiments.
In that experiments, we did not check the invertible property of the randomly generated
$2\times 2$ matrices $A_0, \ldots, A_{559}$. Thus the previously reported experimental
results are not accurate. The experimental evidence confirms our expectatiohn that RLCE scheme behaves like a random linear code.

\subsection{Algebraic attacks}
Faugere, Otmani, Perret, and Tillich \cite{fau2010algebraicae} developed an
algebraic attack against quasi-cyclic and dyadic structure based 
compact variants of McEliece encryption scheme. In a high level,
the algebraic attack from \cite{fau2010algebraicae} tries 
to find ${\bf x^*,y^*}\in GF(q)^n$ such that $V_t({\bf x^*,y^*})$
is the parity check matrix for the underlying alternant codes
of the  compact variants of McEliece encryption scheme. $V_t({\bf x^*,y^*})$
can then be used to break the McEliece scheme.
Note that this $V_t({\bf x^*,y^*})$ is generally different from the original 
parity check matrix $V_t({\bf x,y})$ in (\ref{parityhv}). 
The parity check matrix $V_t({\bf x^*,y^*})$ was obtained  by 
solving an equation system constructed from
\begin{equation}
\label{vteqte}
V_t({\bf x^*,y^*}){G}^T=0,
\end{equation}
where $G$ is the public key.  The authors of \cite{fau2010algebraicae} 
employed the special properties of quasi-cyclic and dyadic structures
(which provide additional linear equations) to rewrite the equation
system obtained from (\ref{vteqte}) and then calculate 
$V_t({\bf x^*,y^*})$ efficiently.

Faugere, Gauthier-Uma\~na, Otmani, Perret, and Tillich \cite{faugere2013distinguisher}
used the algebraic attack in \cite{fau2010algebraicae}  to design an efficient
Goppa code distinguisher to distinguish a random matrix from 
the matrix of a Goppa code whose rate is close to 1. For instance,  
\cite{faugere2013distinguisher} showed that the binary Goppa code 
obtained with $m = 13$ and $r = 19$ 
corresponding to a $90$-bit security key is distinguishable.

It is challenging to mount the above mentioned algebraic attacks  
on the RLCE encryption scheme. Assume that the RLCE scheme is based on Reed-Solomon code. Let 
$G$ be the public key and $(S,G_s, A,P)$ be the private key.
The parity check matrix for a Reed-Solomon code is in the format of 
\begin{equation}
\label{valsf}
V_t(\alpha)=\left[
\begin{array}{ccccc}
1 & \alpha & \alpha^2 &\cdots & \alpha^{n-1}\\
1&\alpha^2&\alpha^4&\cdots&\alpha^{2(n-1)}\\
\vdots &\vdots &\vdots &\ddots &\vdots \\
1&\alpha^{t+1}&\alpha^{2(t+1)}&\cdots&\alpha^{(t+1)(n-1)}\\
\end{array}
\right].
\end{equation}
The algebraic attack in \cite{faugere2013distinguisher,fau2010algebraicae} 
requires one to obtain a parity check matrix $V_t(\alpha^*)$ for the underlying 
Reed-Solomon code from the public key $G$, where $\alpha^*$ may be
different from $\alpha$. Assume that $V_t(\alpha^*)=[{\bf v}_0, \cdots, {\bf v_{n-1}}]\in GF(q)^{(t+1)\times n}$
is a parity check matrix for the underlying Reed-Solomon code.
Let $V'_t(\alpha^*)\in GF(q)^{(t+1)\times n(r+1)}$ be a ${(t+1)\times n(r+1)}$ matrix
obtained from $V_t(\alpha^*)$ by inserting $r$ column vectors ${\bf 0}$ 
after each column of $V_t(\alpha^*)$. That is,
\begin{equation}
\label{fafrebfg}
V'_t(\alpha^*)=[{\bf v}_0, {\bf 0}, {\bf v}_1, {\bf 0}, \cdots, {\bf v_{n-1}}, {\bf 0}].
\end{equation}
Then we have 
\begin{equation}
\label{farer}
\begin{array}{lll}
V'_t(\alpha^*)G_1^T&=& V'_t(\alpha^*)
[{\bf g}_0, C_0, \cdots,  {\bf g}_{n-1}, C_{n-1}]^T\\
&=&V_t(\alpha^*) [{\bf g}_0, \cdots,  {\bf g}_{n-1}]^T\\
&=&V_t(\alpha^*)G_s^T\\
&=&{\bf 0}.
\end{array}
\end{equation}

We cannot build an equation system for the unknown $V'_t(\alpha^*)$ from the public
key $G=SG_1AP$ directly since the identity (\ref{farer}) only shows the relationship
between $V'_t(\alpha^*)$ and $G_1$. In other words, in order to build an equation system
for $V'_t(\alpha^*)$, one also needs to use unknown variables 
for the non-singular matrix $A$ and the permutation matrix $P$. That is, we have 
\begin{equation}
\label{rlceveqaf}
V'_t(\alpha^*)(A^{-1})^T(P^{-1})^TG^T=V'_t(\alpha^*)(GP^{-1}A^{-1})^T
=V'_t(\alpha^*)G_1^TS^T={\bf 0}.
\end{equation}
with an unknown $\alpha^*$, an  unknown  permutation matrix $P$,  and an unknown matrix 
$A=\mbox{diag}[A_0, \cdots, A_{n-1}]$ which consists of $n$ dense nonsingular 
$(r+1)\times (r+1)$ matrices $A_i\in GF(q)^{(r+1)\times (r+1)}$ as defined in (\ref{defineA}).
In order to find a solution $\alpha^*$, one first needs to 
take a potential permutation matrix $P^{-1}$  to reorganize columns of the public key $G$.
Then, using the identity $V'_t(\alpha^*)(A^{-1})^T(P^{-1})^TG^T={\bf 0}$, 
one can build a degree $(t+1)(n-1)+1$ equation system  of $k(t+1)$ equations in $n(r+1)^2+1$ unknowns.
In case that $k(t+1)\ge n(r+1)^2+1$, one may use 
Buchberger's Gr\"{o}bner basis algorithms as in  \cite{fau2010algebraicae} 
to find a solution $\alpha^*$. However, this kind of algebraic attacks are infeasible due to the following 
two challenges. First  the number of permutation matrices $P$ is too large to be handled practically. 
Secondly, even if one can manage to handle the 
large number of permutation matrices $P$, the Gr\"{o}bner basis 
(or the  improved variants such as  F$_4$
or F$_5$ in Faugere \cite{faugere1999new,faugerenewf5}) are impractical 
for such kind of equation systems.

The Gr\"{o}bner basis algorithm eliminates top order monomial (in a given 
order such as lexicographic order) by combining two equations 
with appropriate coefficients. This process continues until one 
obtains a univariate polynomial equation.  
The resulting univariate polynomial equation normally has 
a very high degree and Buchberger's algorithm runs 
in exponential time on average
(the worst case complexity is double exponential time).
Thus Buchberger's algorithm cannot solve nonlinear multicariate equation
systems with more than $20$ variables in practice 
(see, e.g., Courtois et al \cite{courtois2000eff}).
But it should also be noted that though the worst-case 
Gr\"{o}bner basis algorithm is double exponential, the generic
behavior is generally much better. In particular, if the algebraic system
has only a finite number of common zeros at infinity, then 
Gr\"{o}bner basis algorithm for any ordering stops in a polynomial time
in $d^n$ where $d=\max\{d_i: d_i\mbox{ is the total degree of }f_i\}$ 
and $n$ is the number of variables (see, e.g., \cite{bardet2004complexity}). 

\section{Practical considerations}
\label{pracsec}
In order to reduce the message expansion ratio which is defined as the rate of 
ciphertext size and corresponding plaintext size, it  is preferred to use a smaller $r$ for 
the RLCE encryption scheme. Indeed, the experimental 
results show that $r=1$ is sufficient for  RLCE to behave like a random linear code.
As mentioned in the introduction section, the most powerful message 
recovery attack (not private key recovery attack) on McEliece encryption 
schemes is the information-set decoding attack. For the RLCE encryption
scheme, the information-set decoding attack is based on the number of
columns in the public key $G$ instead of the number of columns in the private key $G_s$.
For the same error weight $t$, the probability to find error-free coordinates 
in $(r+1)n$ coordinates is different from the probability to find error-free coordinates
in $n$ coordinates.  Specifically,  the cost of information-set decoding attacks on 
an $[n,k, t;r]$-RLCE scheme is equivalent to the cost of  information-set decoding attacks on a standard 
$[(r+1)n,k; t]$-McEliece scheme.

Taking into  account of the cost of recovering McEliece encryption scheme secret keys 
from the public keys  and 
the cost of recovering McEliece encryption scheme plaintext messages
from ciphertexts using the information-set decoding methods, we generated
a  recommended list of parameters for RLCE scheme in Table \ref{lfarr} 
using the PARI/GP script by Peters's \cite{peters2010information}.
For the recommended parameters, the default underlying linear code is taken as
the Reed-Solomon code over $GF(q)$  and the value of $r$ is taken as $1$.
For the purpose of comparison, we also list the recommended parameters from \cite{bernstein2008attacking} 
for binary Goppa code based McEliece encryption scheme.
The authors in \cite{bernstein2008attacking,peters2010information} proposed the 
use of semantic secure message coding approach so that one can store 
the public key as a systematic generator matrix. For 
binary Goppa code based McEliece encryption scheme, the systematic generator matrix
public key is $k(n-k)$ bits. For RLCE encryption scheme over $GF(q)$,
the systematic generator matrix public key is $k(n(r+1)-k)\log q$ bits.
It is observed that RLCE scheme generally has larger but acceptable public key size.  
Specifically, for the same security level,
the public key size for the RLCE scheme is approximately four to  five times larger than 
the public key size for binary Goppa code based McEliece encryption scheme.
For example, for the security level of 80 bits, the binary Goppa code based 
McEliece encryption scheme has a public key of size 
$56.2$KB, and the RLCE-MDS scheme
has a public key of size $267\approx 5\times 56.2$KB.

\begin{table}[h]
\caption{Parameters for RLCE: $n, k, t, q$, key size ($r=1$ for all parameters), where 
``$360,200, 80, 2^{8}, 101$KB'' under column ``RLCE-MDS code'' represents
$n=360, k=200, t=80$, and $q=2^8$.}
\label{lfarr}
\begin{center}
\begin{tabular}{|l|l|l|l|} \hline
Security& RLCE-MDS code &binary Goppa code \cite{bernstein2008attacking} \\ \hline
60 &360,200, 80, $2^{8}$, 101KB & 1024, 524, 50, 19.8KB \\ \hline
80 &560, 380, 90, $2^{8}$, 267KB& 1632, 1269, 34, 56.2KB \\ \hline
128 &1020, 660, 180, $2^{9}$, 0.98MB&2960, 2288, 57, 187.7KB\\ \hline
192 &1560, 954, 203, $2^{10}$, 2.46MB&4624, 3468, 97, 489.4KB\\ \hline
256 &2184, 1260, 412, $2^{10}$, 4.88MB&6624, 5129, 117, 0.9MB \\ \hline 
\end{tabular}
\end{center}
\end{table}

\section{Conclusions}
In this paper, we presented techniques for designing general random linear code 
based public encryption schemes using any linear code. Heuristics and experiments 
encourages us to think that the proposed  schemes are immune against existing 
attacks on linear code based encryption schemes such as 
Sidelnikov-Shestakov attack, filtration attacks,
and algebraic attacks.
In addition to being a post-quantum cryptographic technique, our scheme RLCE has recently
been used by Wang and Desmedt \cite{rlcehomo} to design fully homomorphic encryption schemes.

\section*{Acknowledgments}
I would like to thank several colleagues for very detailed comments and suggestionsto improve the presentation of this paper.

\bibliographystyle{plain}


\begin{thebibliography}{10}

\bibitem{baldi2008new}
M.~Baldi, M.~Bodrato, and F.~Chiaraluce.
\newblock A new analysis of the mceliece cryptosystem based on qc-ldpc codes.
\newblock In {\em Security and Cryptography for Networks}, pages 246--262.
  Springer, 2008.

\bibitem{bardet2004complexity}
M.~Bardet, J.-C. Faugere, and B.~Salvy.
\newblock On the complexity of {G}r{\"o}bner basis computation of semi-regular
  overdetermined algebraic equations.
\newblock In {\em Proc. Int. Conference on Polynomial System Solving}, pages
  71--74, 2004.

\bibitem{becker2012decoding}
A.~Becker, A.~Joux, A.~May, and A.~Meurer.
\newblock Decoding random binary linear codes in $2^{n/20}$: How $1+1=0$
  improves information set decoding.
\newblock In {\em EUROCRYPT 2012}, pages 520--536. Springer, 2012.

\bibitem{berger2009reducing}
T.P. Berger, P.-L. Cayrel, P.~Gaborit, and A.~Otma\~{n}i.
\newblock Reducing key length of the mceliece cryptosystem.
\newblock In {\em Progress in Cryptology--AFRICACRYPT 2009}, pages 77--97.
  Springer, 2009.

\bibitem{berger2005mask}
T.P Berger and P.~Loidreau.
\newblock How to mask the structure of codes for a cryptographic use.
\newblock {\em Designs, Codes and Cryptography}, 35(1):63--79, 2005.

\bibitem{bernstein2011wild}
D.~Bernstein, T.~Lange, and C.~Peters.
\newblock Wild {McEeliece}.
\newblock In {\em Selected Areas in Cryptography}, pages 143--158. Springer,
  2011.

\bibitem{bernstein2008attacking}
D.J. Bernstein, T.~Lange, and C.~Peters.
\newblock Attacking and defending the {McEliece} cryptosystem.
\newblock In {\em Post-Quantum Cryptography}, pages 31--46. Springer, 2008.

\bibitem{courtois2000eff}
N.~Courtois, A.~Klimov, J.~Patarin, and A.~Shamir.
\newblock Efficient algorithms for solving overdefined systems of multivariate
  polynomial equations.
\newblock In {\em EUROCRYPT 2000}, pages 392--407. Springer, 2000.

\bibitem{couvreur2013distinguisher}
A.~Couvreur, P.~Gaborit, V.~Gauthier-Uma{\~n}a, A.~Otmani, and J.-P. Tillich.
\newblock Distinguisher-based attacks on public-key cryptosystems using
  {Reed--Solomon} codes.
\newblock {\em Designs, Codes and Cryptography}, pages 1--26, 2013.

\bibitem{couvreur2014isit}
A.~Couvreur, I.~M{\'a}rquez-Corbella, and R.~Pellikaan.
\newblock A polynomial time attack against algebraic geometry code based public
  key cryptosystems.
\newblock In {\em Proc. ISIT}, pages 1446--1450. IEEE, 2014.

\bibitem{couvreur2014polynomial}
A.~Couvreur, A.~Otma\~{n}i, and J.-P. Tillich.
\newblock Polynomial time attack on wild {McEliece} over quadratic extensions.
\newblock In {\em Advances in Cryptology--EUROCRYPT 2014}, pages 17--39.
  Springer, 2014.

\bibitem{faugerenewf5}
J.-C. Faugere.
\newblock A new efficient algorithm for computing {G}r{\"o}bner bases without
  reduction to 0 ({F5}).
\newblock In {\em Proc. ISSAC}, pages 75--83.

\bibitem{faugere1999new}
J.-C. Faugere.
\newblock A new efficient algorithm for computing {G}r{\"o}bner bases ({F4}).
\newblock {\em J. Pure and Applied Algebra}, 139(1):61--88, 1999.

\bibitem{faugere2013distinguisher}
J.-C. Faugere, V.~Gauthier-Uma\~{n}a, A.~Otmani, L.~Perret, and J.-P. Tillich.
\newblock A distinguisher for high-rate mceliece cryptosystems.
\newblock {\em Information Theory, IEEE Transactions on}, 59(10):6830--6844,
  2013.

\bibitem{fau2010algebraicae}
J.-C. Faugere, A.~Otmani, L.~Perret, and J.-P. Tillich.
\newblock Algebraic cryptanalysis of {McEliece} variants with compact keys.
\newblock In {\em Eurocrypt 2010}, pages 279--298. Springer, 2010.

\bibitem{faugere2014algebraic}
J.-C. Faugere, L.~Perret, and F.~De~Portzamparc.
\newblock Algebraic attack against variants of mceliece with goppa polynomial
  of a special form.
\newblock In {\em Advances in Cryptology--ASIACRYPT 2014}, pages 21--41.
  Springer, 2014.

\bibitem{janwa1996mceliece}
H.~Janwa and O.~Moreno.
\newblock Mceliece public key cryptosystems using algebraic-geometric codes.
\newblock {\em Designs, Codes and Cryptography}, 8(3):293--307, 1996.

\bibitem{lee1988observation}
P.~J. Lee and E.~F. Brickell.
\newblock An observation on the security of {McEliece's} public-key
  cryptosystem.
\newblock In {\em EUROCRYPT'88}, pages 275--280. Springer, 1988.

\bibitem{leon1988probabilistic}
J.~Leon.
\newblock A probabilistic algorithm for computing minimum weights of large
  error-correcting codes.
\newblock {\em IEEE Trans. Information Theory}, 34(5):1354--1359, 1988.

\bibitem{londahl2012new}
C.~L{\"o}ndahl and T.~Johansson.
\newblock A new version of mceliece pkc based on convolutional codes.
\newblock In {\em Information and Communications Security}, pages 461--470.
  Springer, 2012.

\bibitem{marquez2012error}
I.~M{\'a}rquez-Corbella and R.~Pellikaan.
\newblock Error-correcting pairs for a public-key cryptosystem.
\newblock {\em arXiv preprint arXiv:1205.3647}, 2012.

\bibitem{may2011decoding}
A.~May, A.~Meurer, and E.~Thomae.
\newblock Decoding random linear codes in $\tilde{O}(2^{0.054 n}$).
\newblock In {\em ASIACRYPT 2011}, pages 107--124. Springer, 2011.

\bibitem{may2015computing}
A.~May and I.~Ozerov.
\newblock On computing nearest neighbors with applications to decoding of
  binary linear codes.
\newblock In {\em EUROCRYPT 2015}, pages 203--228. Springer, 2015.

\bibitem{mceliece1978public}
R.J. McEliece.
\newblock A public-key cryptosystem based on algebraic coding theory.
\newblock {\em DSN progress report}, 42(44):114--116, 1978.

\bibitem{misoczki2009compact}
R.~Misoczki and P.~Barreto.
\newblock Compact mceliece keys from goppa codes.
\newblock In {\em Selected Areas in Cryptography}, pages 376--392. Springer,
  2009.

\bibitem{misoczki2013mdpc}
R.~Misoczki, J.-P. Tillich, N.~Sendrier, and P.~Barreto.
\newblock {MDPC-McEliece}: New {McEliece} variants from moderate density
  parity-check codes.
\newblock In {\em Information Theory Proceedings (ISIT), 2013 IEEE
  International Symposium on}, pages 2069--2073. IEEE, 2013.

\bibitem{niederreiter1986knapsack}
H.~Niederreiter.
\newblock Knapsack-type cryptosystems and algebraic coding theory.
\newblock {\em Problems of Control and Information Theory}, 15(2):159--166,
  1986.

\bibitem{peters2010information}
C.~Peters.
\newblock Information-set decoding for linear codes over ${F}_q$.
\newblock In {\em Post-Quantum Cryptography}, pages 81--94. Springer, 2010.

\bibitem{prange1962use}
Eugene Prange.
\newblock The use of information sets in decoding cyclic codes.
\newblock {\em IRE Trans. Information Theory}, 8(5):5--9, 1962.

\bibitem{shoup2001ntl}
Victor Shoup.
\newblock {NTL}: A library for doing number theory, 2001.

\bibitem{sidelnikov1992insecurity}
V.~M Sidelnikov and S.~Shestakov.
\newblock On insecurity of cryptosystems based on generalized {Reed-Solomon}
  codes.
\newblock {\em Discrete Mathematics and Applications}, 2(4):439--444, 1992.

\bibitem{sidelnikov1994public}
V.M. Sidelnikov.
\newblock A public-key cryptosystem based on binary reed-muller codes.
\newblock {\em Discrete Mathematics and Applications}, 4(3):191--208, 1994.

\bibitem{stern1989method}
J.~Stern.
\newblock A method for finding codewords of small weight.
\newblock In {\em Coding theory and applications}, pages 106--113. Springer,
  1989.

\bibitem{rlcehomo}
Y.~Wang and Y.~Desmedt.
\newblock Towards fully homomorphic encryption schemes from codes.
\newblock In {\em Submitted}, pages 1--1. Springer Press, 2015.

\bibitem{wangrlcesoft}
Yongge Wang.
\newblock {RLCE} implementation \url{http://webpages.uncc.edu/yonwang/rlce},
  2015.

\bibitem{wieschebrinkreisedMce}
C.~Wieschebrink.
\newblock Two {NP}-complete problems in coding theory with an application in
  code based cryptography.
\newblock In {\em Proc. IEEE ISIT}, pages 1733--1737. IEEE Press, 2006.

\bibitem{wieschebrink2010cryptanalysis}
C.~Wieschebrink.
\newblock Cryptanalysis of the {Niederreiter} public key scheme based on {GRS}
  subcodes.
\newblock In {\em Post-Quantum Cryptography}, pages 61--72. Springer, 2010.

\end{thebibliography}

\end{document}